\documentclass[%
reprint,
 amsmath,amssymb,
 aps,
 prl,
]{revtex4-2}

\usepackage{graphicx}
\usepackage{dcolumn}
\usepackage{bm}
\usepackage{xcolor}

\usepackage[utf8]{inputenc}
\usepackage{amsmath}
\usepackage{amsfonts}
\usepackage{amssymb}
\usepackage{hyperref}
\usepackage{subcaption}
\usepackage[normalem]{ulem}
\usepackage[percent]{overpic}

\newcommand{\Lcal}{\mathcal{L} }
\newcommand{\Ical}{\mathcal{I} }

\newcommand{\Ocal}{\mathcal{O} }

\newcommand{\rr}{\mathbf{r} }
\newcommand{\vv}{\mathbf{v} }
\newcommand{\qq}{\mathbf{q} }
\newcommand{\FF}{\mathbf{F} }

\newcommand{\RR}{\mathbf{R} }
\newcommand{\KK}{\mathbf{K} }
\newcommand{\uu}{\mathbf{u}}

\newcommand{\ff}{\mathbf{f}}

\newcommand{\qqhat}{\hat{\mathbf{q}} }

\newcommand{\xxhat}{\hat{\mathbf{x}} }

\newcommand{\zzhat}{\hat{\mathbf{z}} }
\newcommand{\pphat}{\hat{\mathbf{p}} }

\newcommand{\utilde}{\tilde{u} }

\begin{document}


\title{Hydrodynamic Mechanism of Colloidal Propulsion through Momentum Exchange }

\author{Javier Diaz}
\affiliation{Departament de F\'isica de la Mat\`eria Condensada, Universitat de Barcelona, Mart\'i i Franqu\'es 1, 08028 Barcelona, Spain}
\affiliation{Universitat de Barcelona Institute of Complex Systems (UBICS), Universitat de Barcelona, 08028 Barcelona, Spain}

\author{Ignacio Pagonabarraga}
\affiliation{Departament de F\'isica de la Mat\`eria Condensada, Universitat de Barcelona, Mart\'i i Franqu\'es 1, 08028 Barcelona, Spain}
\affiliation{Universitat de Barcelona Institute of Complex Systems (UBICS), Universitat de Barcelona, 08028 Barcelona, Spain}

\author{Carles Calero}
\affiliation{Departament de F\'isica de la Mat\`eria Condensada, Universitat de Barcelona, Mart\'i i Franqu\'es 1, 08028 Barcelona, Spain}
\affiliation{Universitat de Barcelona Institute of Complex Systems (UBICS), Universitat de Barcelona, 08028 Barcelona, Spain}


\begin{abstract}

Propulsion of colloidal particles due to momentum transfer from localized surface reactions is investigated by solving the exact unsteady Stokes equation. 
We model the effect of surface reactions as either a {\it force dipole} acting on the fluid or a {\it pair force} acting on both the colloid and the fluid. 
Our analysis reveals that after a single reaction event the colloid's velocity initially decays as $\sim t^{-1/2}$, followed by a long-time tail decay $\sim t^{-5/2}$. This behavior is distinct from the $\sim t^{-3/2}$ decay seen for simple impulsively forced particles, a result of the force-free nature of the reaction mechanism. 
The velocity and transient dynamics are strongly controlled by the distance of the reaction from the colloid surface. 
For a colloid subject to periodic reactions, the theory predicts a steady-state velocity that is comparable to experimental results and previous simulations, suggesting that direct momentum transfer is a relevant mechanism for self-propulsion in systems like Janus particles. 
Finally, our study shows that fluid compressibility is not required for momentum transfer to produce colloidal propulsion.

\end{abstract}

\maketitle

Particle self-propulsion leads to non-equilibrium collective dynamics which, in the last decades, has seen a considerable range of emergent and complex phenomena\cite{vicsek_novel_1995,fily_athermal_2012,sanchez_spontaneous_2012,redner_structure_2013,digregorio_full_2018}. 
Micron-sized synthetic particles often rely on self-phoretic mechanisms to sustain their motion\cite{anderson_colloid_1989}, that is, the colloid itself generates a gradient of, \textit{e.g.}, temperature, concentration or electric field, respectively called (self-) 
thermophoretic\cite{jiang_active_2010,yang_simulations_2011}, 
diffusophoretic\cite{paxton_catalytic_2004,golestanian_propulsion_2005} or 
electrophoretic particles\cite{wang_bipolar_2006,gibbs_asymmetric_2010,wheat_rapid_2010}. 
These self-propelled synthetic colloids have been realized experimentally with considerable success. 
For their potential applications to be realized, there is an effort to synthesize increasingly smaller self-propelled particles~\cite{golestanian_enhanced_2015,bechinger_active_2016,calero_self-thermophoresis_2020,arque_enzyme-powered_2022,jurado_romero_high_2023}.

Colloidal propulsion often occurs through {\it interfacial reactions} that are asymmetrically distributed on their surface, such as the case of Janus particles that catalyze chemical reactions on one of their hemispheres or the case of enzymes~\cite{Riedel_Nature2015, golestanian_enhanced_2015}.
Microparticles partially coated with Pt in a $H_2O_2$ solution~\cite{howse_self-motile_2007} are a typical example, which propel due to the exothermic reaction at the Pt surface; however, the origin of propulsion in these systems has been disputed for a long time~\cite{wang_open_2023,bishop_active_2023,lyu_active_2021}, with early experiments suggesting self-diffusophoresis of neutral solutes~\cite{zhang_self-propulsion_2017}, while later  works have identified the role of salts, suggesting ionic self-diffusophoresis~\cite{brown_ionic_2014} or self-electrophoretic effects~\cite{lyu_active_2021}. Another interfacial mechanism responsible for self-propulsion consists of the asymmetric bubble formation and release at the microparticles' surface, observed for different geometries and shown to produce large speeds and persistent dynamics~\cite{wang_Langmuir2014, solovev_ACSNanoc2012, gao_JACS2011}.
Further, short-lived interfacial solute-solvent forces are thought to be responsible for propulsion in a recent proposal to downsize active matter to the nanoscale~\cite{calero_self-thermophoresis_2020, jurado_romero_high_2023}. In these numerical studies, well-known mechanisms of molecular excitation with external radiation and energy relaxation into the solvent~\cite{Yardley1980, Rey2009, Rey2015} are employed to generate directed motion of asymmetric nanoparticles.

In these processes, {\it reactions} occur following a time-dependent distribution, which can be stochastic, as in the case of chemical reactions (in Janus particles or enzymes) or bubble formation~\cite{wang_Langmuir2014,solovev_ACSNanoc2012}, or periodic as in cases with external actuation or cyclical reshaping of swimmers. 
The reaction kinetics  is characterized by the average reaction rate  in the former case, and by an actuation frequency  in the latter, which define the characteristic time-scale of the process. 
These time-scales can couple to the hydrodynamic time-scales, determining the dynamics of the propeller. For an incompressible liquid of kinematic viscosity $\nu$, momentum impact on a colloid of radius $a$ develops over a time scale $t_a=a^2/\nu$, which ranges from pico to microseconds for colloids between the nano and micro scales. The kinetics of chemical reactions and molecular cycling can vary  over many decades in frequency, and hence can  become  comparable to $t_a$.

The contribution to transport from phoretic effects arising from the emergence of surface gradients has been extensively investigated~\cite{anderson_colloid_1989, golestanian_propulsion_2005}. 
Alternative propulsion mechanisms involving direct momentum transfer from the solvent to the particle have been proposed, but they remain relatively underexplored. The contribution to particle motility of pressure waves produced by chemical reactions in a compressible solvent was studied in Ref.~\cite{felderhof_dynamics_2010}. Recently, the role of direct chemo-mechanical coupling, where the chemical reaction  leads to a direct momentum transfer in the liquid, has been advocated through coarse-grained particle-based simulations~\cite{eloul_reactive_2020}, highlighting  the coupling between the reaction kinetics and the hydrodynamic flows.

In this work, we employ a hydrodynamic theory to investigate the role of momentum transfer from inhomogeneous surface reactions in driving colloidal propulsion. 
To isolate the essential features of the coupling between interfacial reactions and hydrodynamic response, we introduce a minimalist reaction scheme that occurs in the fluid surrounding the colloid and self-consistently generates the resulting flow field. 
Our analysis is restricted to colloidal particles operating at low Reynolds numbers, although we do not assume creeping flow, as capturing the time-dependent response of the fluid to reaction kinetics is essential. 
We assume fluid incompressibility in contrast to previous approaches~\cite{felderhof_dynamics_2010,eloul_reactive_2020}, which will allow us to assert the necessity of compressibility in generating propulsion.   
This description, based on unsteady Stokes dynamics, overcomes particle-size limitations  encountered in particle-based simulations.

\begin{figure}
    \centering
    \includegraphics[width=1\linewidth]{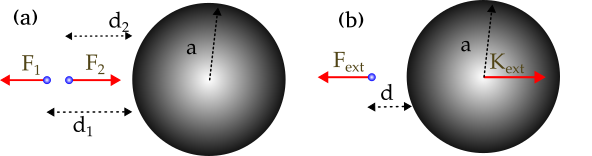}
    \caption{
    Schematic view of the two scenarios considered in this work for a colloid with radius $a$: 
       (a) a dipole of point force densities act on the fluid at distances $d_1$ and $d_2$ from the colloidal surface; 
       (b) a pair of forces with $\KK_{ext}$ acting directly on the colloid and an opposite force $\FF_{ext}$ on the fluid at a distance $d$ from the surface. 
    }
    \label{fig:scheme}
\end{figure}

We consider a colloidal sphere of radius $a$, density $\rho_p$, and mass $m=4\pi a^3\rho_p/3$ immersed in an incompressible fluid with density $\rho$ and kinematic viscosity $\nu=\eta/\rho$.  
In our model, the essential effect of an interfacial reaction is described as a mechanical pulse on the fluid, through a force- and torque-free instantaneous transfer of momentum. Two possible mechanisms for momentum transfer are considered. 
In Fig.~\ref{fig:scheme}(a), the interfacial reaction is represented as a \textit{force dipole} acting on the fluid, corresponding to the case where reactants and products remain in solution and the colloid merely catalyzes the reaction in its vicinity. In Fig.~\ref{fig:scheme}(b), we adopt the \textit{pair force} scenario, where the reaction occurs at a complex on the colloidal surface. Here, momentum transfer is modeled as equal and opposite forces acting on the fluid and on the colloid’s center of mass. 
This scenario encompasses chemical reactions where reactants bind to the colloid’s surface and products are released, as well as cases where externally excited nanoparticles dissipate excess energy anisotropically into the surrounding solvent~\cite{calero_self-thermophoresis_2020, jurado_romero_high_2023}.
In the force dipole case (Fig.~\ref{fig:scheme}(a)), the  two opposing force densities are located at $\RR_1=-(a+d_1)\zzhat$ and $\RR_2=-(a+d_2)\zzhat$, with the positive force density located closer to the colloid at a distance $d_2$ from its surface, while the negative one is further from it, at a distance $d_1>d_2$ from the surface. 
This defines a separation distance $\delta=d_1-d_2$ and the displacement of the dipole from the colloidal surface as $d=(d_1+d_2)/2$.  
In the pair force scenario (Fig.~\ref{fig:scheme}(b)),  a single force density acting on the fluid is considered, located at a distance $d$ from the colloidal surface at a point in space $\RR=-(a+d)\zzhat$.
Additionally, a force $\KK_{ext}$ acts on the center of mass of the colloid. 
Throughout this work, we will use primed notation to denote magnitudes scaled with the colloidal radius: 
$d_1^{\prime}\equiv d_1/a$, $\delta'\equiv \delta/a$, etc.

The colloid may be subjected to both an unsteady external force applied to its center of mass, $\KK_{ext}(t)$, and a force mediated by the fluid, $\KK(t)$, due to the interfacial reaction. 
The colloidal velocity, $\uu(t)$, then satisfies 
\begin{equation}
    m\frac{d\uu(t)}{dt} =\KK_{ext}(t)+\KK(t).
    \label{eq:newton}
\end{equation}
where the fluid-mediated force $\KK(t)$ depends both on the colloidal velocity $\uu(t)$ and the fluid velocity in the absence of the colloid $\vv_0(\rr,t)$, using the induced forces formalism by Mazur and Bedeaux~\cite{mazur_generalization_1974}. 
Eq.~\ref{eq:newton} in Fourier space then reads
\begin{equation}
    \uu(\omega) = 
    \frac{\KK_{ext}(\omega)/\gamma_0 +(1+\alpha )\overline{\vv}_0^s(\omega) +\frac{1}{3} \alpha^2 \overline{\vv}^v_0(\omega)}{1+\alpha+\beta^{-1}\alpha^2 }
    \label{eq:u-omega}
\end{equation}
where $\alpha =\sqrt{i \omega a^2/\nu}$, $\gamma_0=6\pi\eta a$ is the  drag coefficient, and the density contrast is characterized by the parameter $\beta=9/(1+2\rho_p/\rho)$~\cite{arminski_effect_1979}. 
The fluid flow in the absence of the colloid contributes to the colloidal velocity \textit{via} $\bar{\vv}_0^s(\omega)$ and $\bar{\vv}_0^v(\omega)$, the fluid velocity averaged over the colloidal surface and volume, respectively. 
This unperturbed fluid velocity satisfies the unsteady Stokes equation in the presence of an applied force distribution  $\FF_{ext}(\rr,t)$ acting on the fluid. 
In Fourier space for both space ($\qq$) and time ($\omega$), 
\begin{equation}
    \vv_0(\qq,\omega) = 
    \frac{\Lcal\cdot\FF_{ext}(\qq,\omega) }{i\omega\rho+\eta q^2} 
    \label{eq:v0}
\end{equation}
where the orthogonal projector tensor is $\Lcal = \Ical-\qqhat\qqhat$ with $\Ical$ being the identity tensor and $\qqhat\qqhat$ the dyadic product of the Fourier space unit vector $\qqhat=\qq/|\qq|$.
The spatial dependence of the applied force distribution on the fluid depends on the scenario considered: 
in the force dipole case $\FF_{ext}(\rr,t)=\ff_0(t)\left[ \delta(\rr-\RR_2) - \delta(\rr-\RR_1) \right]$; 
in the pair force case $\FF_{ext}(\rr,t)=-\ff_0(t)\delta(\rr-\RR)$, where we impose  $\KK_{ext}(t)=\ff_0(t)$ to ensure a force-free system.

To capture the features of the motion induced by an interfacial reaction, we first analyze the momentum transfer associated to a single event, of magnitude $\Delta p$, as characterized by a single instantaneous forcing along the $Z$ direction, $\ff_0(t)=\Delta p \delta(t)\zzhat$, which we will later generalize.

\begin{figure}[b]
    \centering
    \begin{overpic}[width=\linewidth]{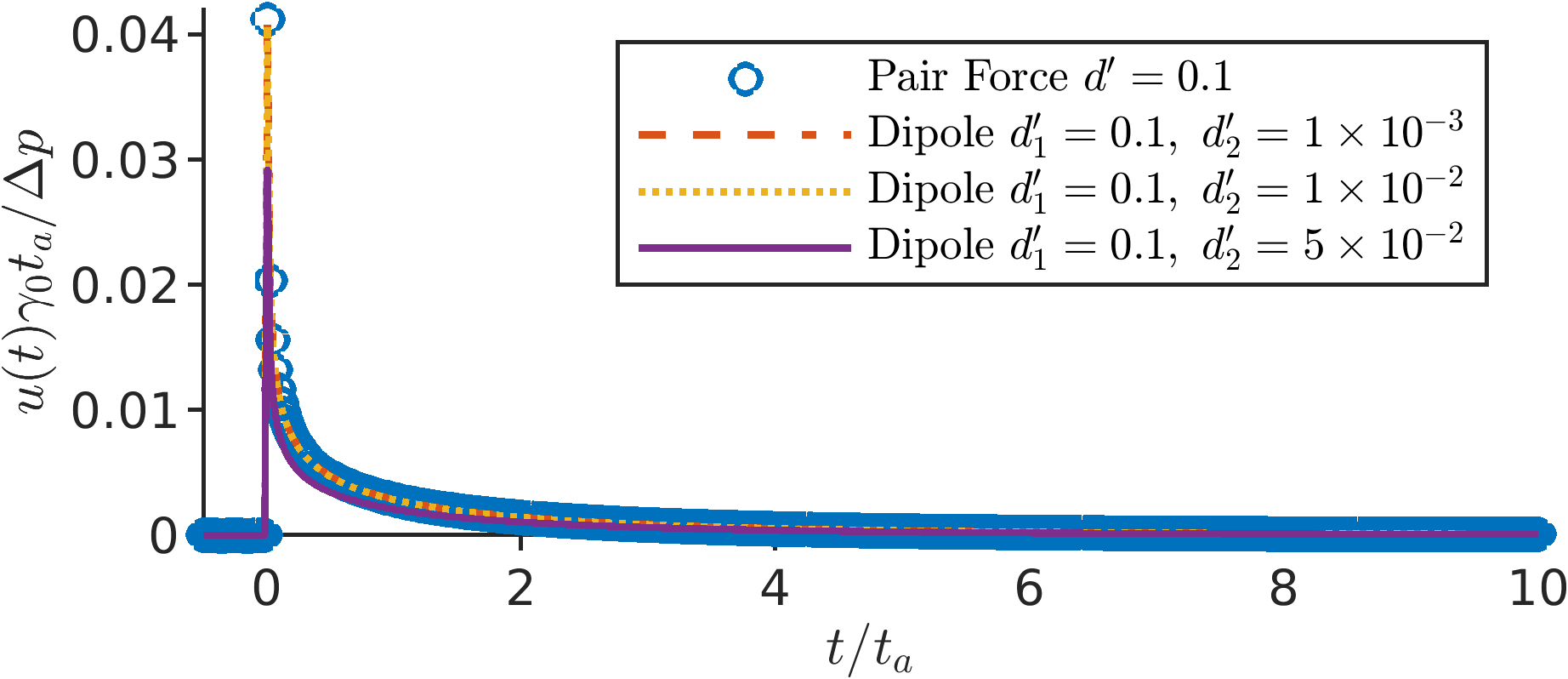}
        \put(3,5){\textbf{(a)}}
    \end{overpic}
    \vspace{1em}
    \begin{overpic}[width=\linewidth]{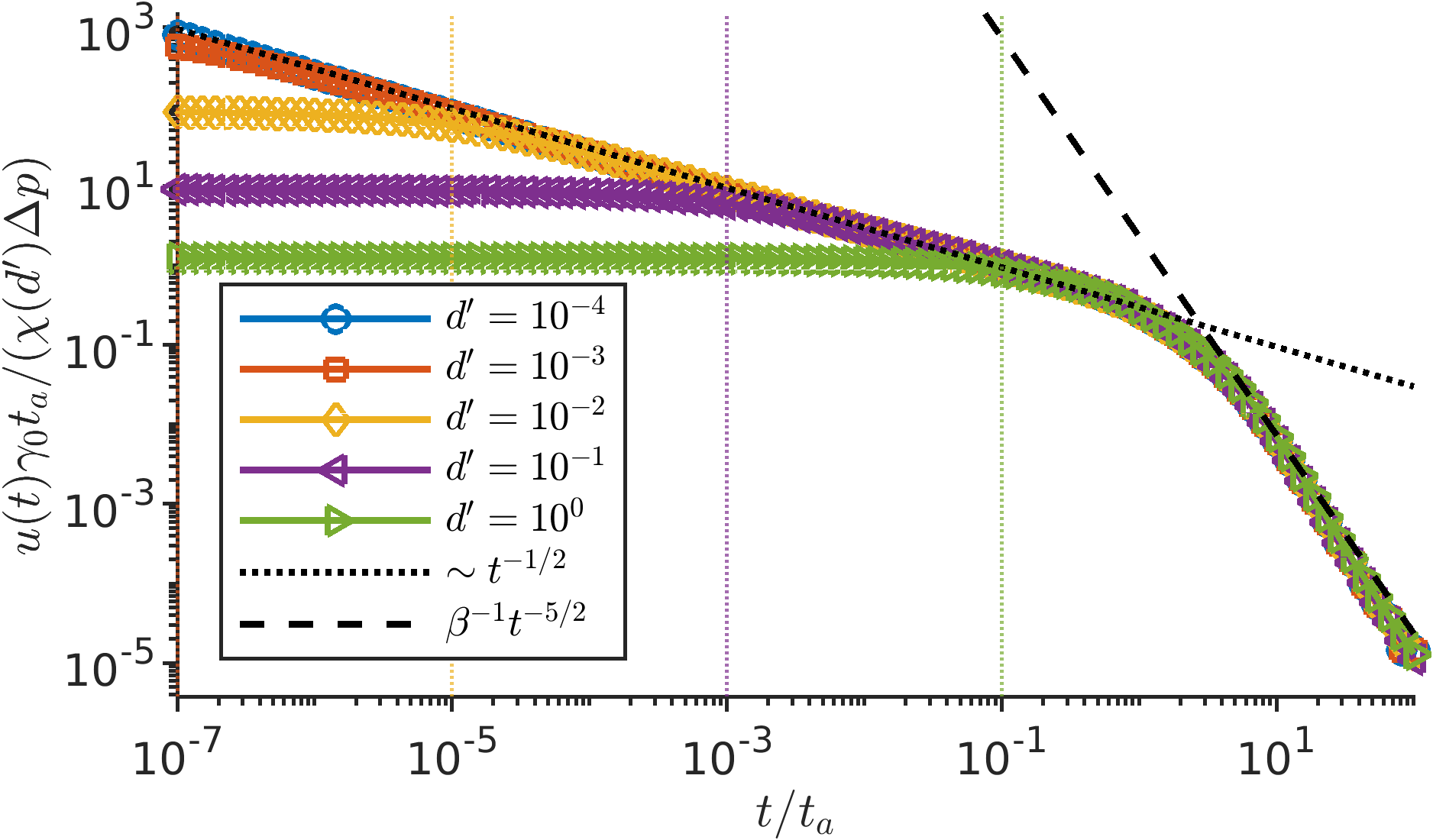}
        \put(3,5){\textbf{(b)}}
    \end{overpic}
        \caption{
        (a) Colloidal velocity following an 
        impulsive pair force with $d^{\prime}=0.1$ and dipole scenario with $d_1^{\prime}=0.1$ and different values of $d_2^{\prime}$. 
        (b) Colloidal velocity on a \textit{log-log} scale for different reaction displacement $d^{\prime}$, scaled with $\chi(d^{\prime})$ (see Eq.~\ref{eq:chid}), along with the mid-time scaling $\sim t^{-1/2}$ and long-time tail prediction $\beta^{-1} t^{-5/2}$. 
        Vertical lines indicate the respective values of $t=t_d/10$ for each value of $d^{\prime}$. 
    }
    \label{fig:ucol}
\end{figure}

Fig.~\ref{fig:ucol}(a) shows the colloidal velocity, $u(t)$, in response to  a single forcing event with $\rho_p/\rho=10$, both for the dipole and the pair force cases. 
In the latter, we consider that the  outer component of the reaction event takes place at a distance $d'=0.1$ from the colloidal surface, while in the former we vary $d_2'$ (the location of the closest reactant to the surface) while keeping $d_1'=0.1$ to quantify the differences between the two scenarios. 
In both cases, the colloid experiences an immediate increase in its velocity due to the impulsive forcing at $t=0$, followed by a monotonic algebraic decay (see below). 
Both scenarios exhibit qualitatively similar behavior, with the two being identical as $d_2\to0$, corresponding to the positive component of the explicit dipole being placed right at the colloidal surface. 
This indicates that the positive component of the momentum transfer in the explicit dipole quickly transfers to the colloidal motion.

The  net displacement of the colloid after a single reaction is given by  $\Delta r=\int_0^{\infty}dt~u(t)=u(\omega=0)$. 
For the pair force scenario, using Eq.~\ref{eq:u-omega} we obtain $\Delta r= \Delta p ~ \chi(d') /\gamma_0 $, with
\begin{equation}
    \chi(d') \equiv  
    \frac{d'^2(2d'+3)}{2(d'+1)^3}
    \label{eq:chid}
\end{equation}
which approximates to $\chi(d') \approx 3d'^2/2$ for small values of $d' \ll1$.
Hence, in the regime  $d'\to0$ the colloidal velocity goes to zero as $\sim 3d'^2/2$, indicating that if the reaction takes place exactly at the colloidal surface, the colloid will not displace.

A closer look at the dynamics of the colloidal velocity in Fig.~\ref{fig:ucol} (b) shows that the displacement $d$ plays a crucial role beyond simply modulating the colloidal speed \textit{via} $\chi(d')$. 
A characteristic time  $t_d\equiv d^2/\nu$ can be defined as the time it takes for the momentum to diffuse over a distance $d$, \textit{i.e.}, for the momentum disturbance due to the reaction to  start affecting  the colloid.

At short times, $0<t<t_d$, $u(t)$ exhibits an extremely slow, barely perceptible decay following the initial momentum transfer.
This initial velocity is proportional to $\beta$, as shown in Fig.~S1 in the SM,  consistent with the behavior of a colloid acted upon only by $\KK_{ext}(t) $\cite{arminski_effect_1979} (\textit{i.e.} in the absence of a reaction). 
The dependence on $\beta$ is explained as an effect of the drag that the motion of the particle exerts on the surrounding fluid. 
Indeed, the acceleration of the particle induces an acceleration of the fluid it displaces, inducing a virtual mass force that augments the particle apparent inertia.

For intermediate times, $t_d<t<t_a$, we observe a regime where $u(t) \sim \chi(d')~t^{-1/2}$ in which both the direct and the fluid-mediated momentum transfers  significantly contribute to the colloidal motion.
The $t^{-1/2}$  dependence can be seen in Fig.~\ref{fig:ucol} (b), encompassing a larger time range for smaller values of $d$.
We can attribute this dependence to the establishment of the diffusive velocity field over the colloidal surface.
This can also be justified analytically as a controlled asymptotic behavior for intermediate frequencies in Fourier space, as shown in Eq.~\ref{eq:u.alpha.intermediate} in Appendix A. 
If the momentum transfer event takes place at a distance comparable to the colloidal size --$d'=1$ in Fig. \ref{fig:ucol} (b)-- then the algebraic dependence $t^{-1/2}$ is no longer present, which resembles the purely impulsive case, shown for reference in Fig.~S2 in the SM with $\FF_{ext}=0$.

For longer times, $t>t_a$, the colloidal velocity exhibits a long-time tail that decays algebraically as $t^{-5/2}$. 
This behavior differs from the well known long-time decay $\sim t^{-3/2}$ of a particle velocity after the action of an impulsive force~\cite{alder_decay_1970,arminski_effect_1979}, as the  dipolar actuation of two opposing point forces  is force-free, leaving only higher order terms, as shown in Eq.~11 in appendix A. 
The algebraic nature of the decay has its origin in the viscous coupling between nonuniform particle motion and unsteady flow generated in the fluid. 
Consequently, this dependence is observed for longer times, such that both the direct and the fluid-mediated contributions to the momentum transfer have diffused over the colloid's length scale.
We also note that the long time velocity scales with $\beta^{-1}$; that is, it is approximately proportional to the colloidal density, which can be found analytically (see Eq.~11 in appendix A) and checked by collapsing $u(t)$ for various values of $\beta$, as in Fig.~S3. 
Overall, the asymptotic long-time  colloidal velocity decays as 
\begin{equation}
    u(t\to\infty) \sim\frac{\Delta p}{\gamma_0} \chi(d')~ \beta^{-1}~(t/t_a)^{-5/2}
\end{equation}
as shown in Fig. \ref{fig:ucol} (b) for $t\gg t_a$. 
In summary, the colloidal velocity exhibits three dynamical regimes controlled by the two diffusion time scales,  $t_d$ and $t_a$. 

Next, we investigate the effect of the reaction kinetics on the colloidal motion.
For simplicity, we describe  the reaction kinetics as a sequence of events equispaced in time, with period $T_0$.  
To understand the relationship between the reaction rate $T_0^{-1}$ and $u(t)$, we model a sequence of periodic reactions leading to a forcing 
$\ff_0(t) =  \zzhat \sum_{k=0}^{\infty} \Delta p~\delta(t-kT_0)$.  
Due to the linearity of the governing equations, the solution for the colloidal velocity can be expressed as
$
    u(t) = \sum_{k=0}^{\infty} u_{(1)} (t-kT_0) 
$
where $u_{(1)}(t)$ is the colloidal velocity under a single forcing at $t=0$ (\textit{i.e.} solving Eq. \ref{eq:u-omega}).

\begin{figure}
    \centering
    \includegraphics[width=1\linewidth]{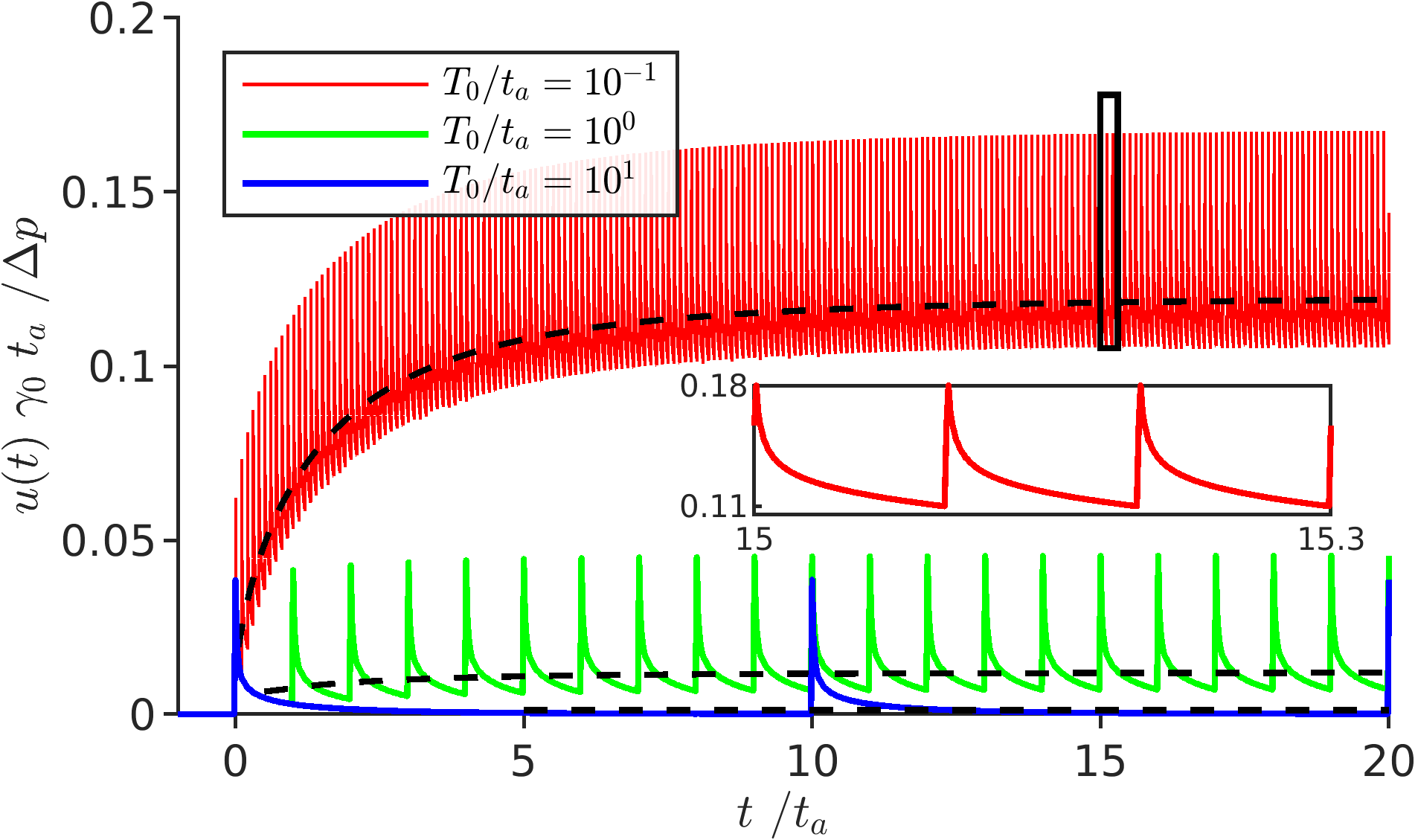}
    \caption{
        Colloidal velocity $u(t)$ under a periodic forcing with three values of the period $T_0$. 
        Black dashed lines for each $T_0$ are the period-averaged velocity $u_n$. 
        The inset is a zoom on the area of the black box within the  $T_0=10^{-1}$ curve.  
    }
    \label{fig:periodic.transient}
\end{figure}

Fig.~\ref{fig:periodic.transient} shows the result for the colloidal velocity $u(t)$  under periodic instantaneous reactions with period $T_0$, characterized by a transient regime followed by a periodic signal around a steady state. 
The average in-period velocity during period $n$ can be calculated as $u_n=T_0^{-1}\int_{nT_0}^{(n+1)T_0}dt~ u(t)$, which is shown along with the instantaneous velocity in Fig.~\ref{fig:periodic.transient}, showing a good agreement. 
Asymptotically,  for $n\gg1$, we obtain an average steady state velocity 
\begin{equation}
    \langle u \rangle_{T_0} = 
    \frac{\Delta p}{\gamma_0 T_0} \chi(d')
    \label{eq:u.steady}
\end{equation}
indicating that a colloid under a periodic forcing reaches an average steady state velocity proportional to $T_0^{-1}$. 
The ratio $\Delta p/T_0$ can be understood as an effective constant force associated with the momentum increment due to the reaction during a period $T_0$. 
In turn, the momentum increase can be related to the energy released during the exothermic reaction $\Delta E$ and the mass of the solute molecules $m_r$, leading to $\Delta p\sim \sqrt{2m_r\Delta E}$. 
Alternatively, we may approximate the periodic actuation on the colloid to the effect of a constant average force $\Delta p/T_0$, which provides equivalent results for the average velocity (see Eq.~E.7 in the SM). Note that, similarly, a periodic application of an instantaneous force dipole would provide results equivalent in the steady state to a constant dipole forcing.   
In the scenario of a typical Pt-Janus colloid in a H$_2$O$_2$ solution and assuming $d/a\sim 0.1$, Eq.~\ref{eq:u.steady} leads to a velocity of the order $\langle u\rangle_{T_0}\sim 5\mu m/s$ which is comparable to the prediction in simulations~\cite{eloul_reactive_2020} and the experimental value~\cite{brown_ionic_2014}.
The similarity between the hydrodynamic prediction and the simulations suggests that fluid compressibility is not required to induce colloidal self-propulsion.

\begin{figure}
    \centering
    \includegraphics[width=1\linewidth]{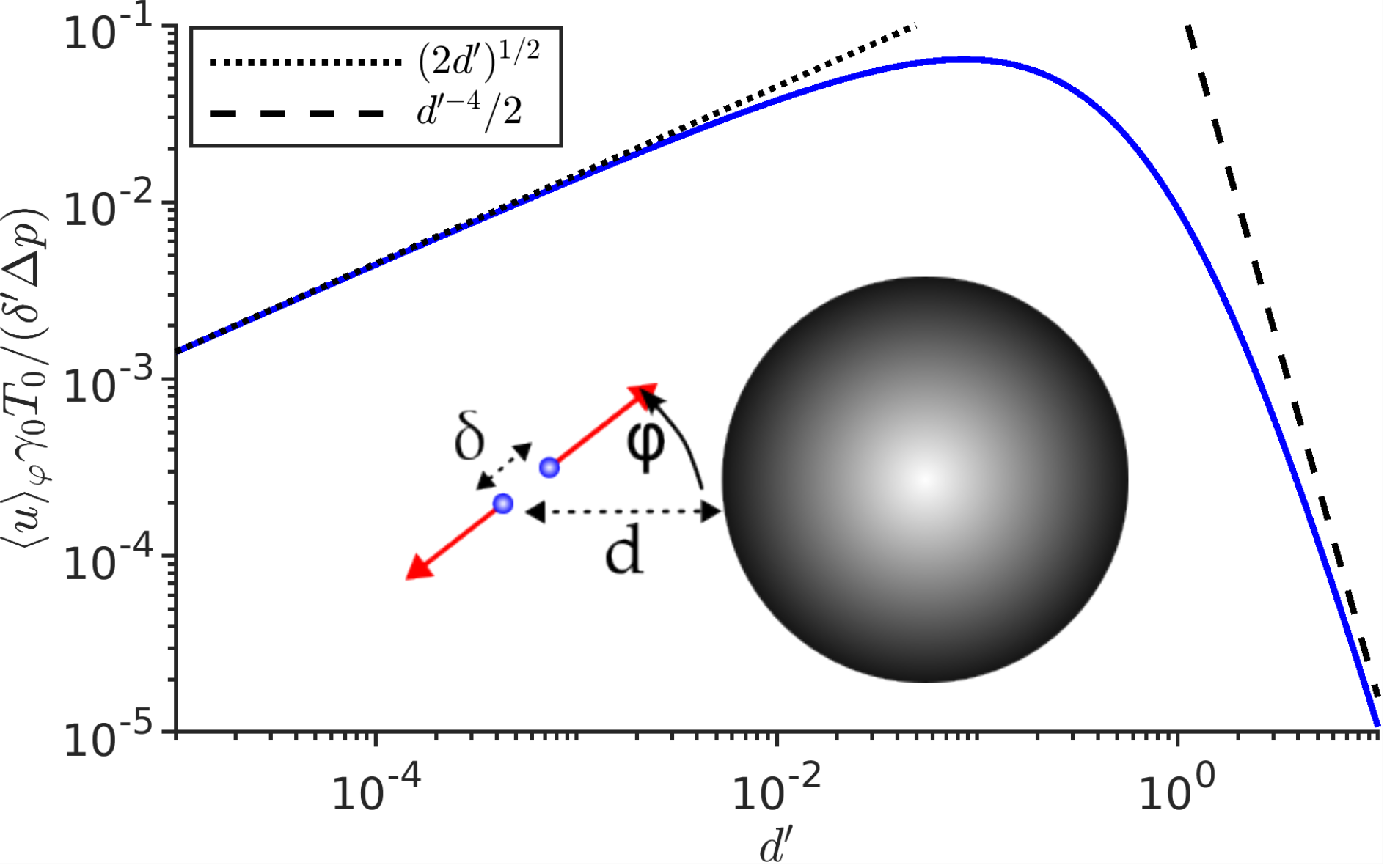}
    \caption{
       Orientationally averaged colloidal velocity $\langle u\rangle_{T_0, \varphi}$ as a function of the dipole displacement $d$ from the colloidal surface. 
    }
    \label{fig:ucol.average.gamma}
\end{figure}

The surface reactions considered so far  assume that the direction of the resulting forces are coaxial with the line connecting to the center of mass of the colloid (see Fig.~\ref{fig:scheme}).
However, a surface reaction might occur along any direction of angle $\varphi$ (see Fig.~\ref{fig:ucol.average.gamma}), with the flow field generated in the $\varphi=0$ case corresponding to a pusher  and $\varphi=\pi/2$ to a puller swimmer.  Using the force dipole model, we can determine the colloid’s average axial velocity arising from periodic instantaneous surface reactions oriented in an arbitrary direction $\varphi$ (see Appendix B).
A particularly interesting case occurs when the reaction is confined along a plane tangent to the colloidal surface (with $\varphi=\pi/2$), which generates a negative average velocity, given by $\langle u (\varphi=\pi/2)\rangle_{T_0} \approx - 3 (\Delta p/\gamma_0T_0 )\delta'd'(d'+2)/( 4(1+d')^4 )$ 
for small $\delta'$ (see complete expression in Eq. E19 and curve agreement in Fig.~S6). Indeed, for $d'>0$ both the numerical and analytical curves indicate a negative colloidal velocity, as the dipole  generates a puller-like fluid flow that attracts the colloid towards the dipole, in agreement with particle-based simulations~\cite{eloul_reactive_2020}.

For a generic isotropic reaction, with a uniform distribution for the reaction angle $\varphi$, we obtain the colloidal mean velocity
\begin{equation}
    \langle u \rangle_{T_0,\varphi} = 
    \frac{\Delta p}{\gamma_0T_0} 
    \frac{\delta'}{\pi}
    \frac{(1+d')\sqrt{d'(d'+2)}-d(2+d')}{(1+d')^4}
     \label{eq:u.mean.gamma}
\end{equation}
which is shown in Fig.~\ref{fig:ucol.average.gamma}, scaled with $\delta'$ in units of $\Delta p/(\gamma_0T_0)$, displaying a maximum at $d'^*\approx 0.08$. 
Contrary to Eq.~\ref{eq:u.steady}, Eq.~\ref{eq:u.mean.gamma} scales linearly with $\delta$, the size associated with the decomposition of the molecule due to the reaction.

In conclusion, a minimal hydrodynamic theory allows us to characterize the self-propulsion of a colloidal particle  where a generic interfacial reaction is modeled as force-free forces acting on the fluid and on the colloid. 
Both short- and long-time behavior exhibit different time dependencies than in the purely impulsive case, with an algebraic long-time tail prediction of $u(t)\sim t^{-5/2}$ and proportional to colloidal density. 
The analytical results suggest that the length scales associated to the local forces play a critical role in the resulting colloidal velocity and directionality, by modulating the average velocity under a periodic forcing, but also determining the short-time behavior. 
The flow considered in our study is incompressible, indicating that momentum transfer through density fluctuations is not necessary for colloids to self-propel.

{\sl Acknowledgments}
We acknowledge support from  the Ministerio de Ciencia, Innovación y Universidades (Spain) Project Nos.  PID2021-126570NB-I00 (I.P.) and PID2021-124297NB-C31 (C.C.); the Spanish grant CNS2022-135395 funded by MCIN/AEI/10.13039/501100011033 and by the European Union NextGenerationEU/PRTR (C.C., J.D.); Generalitat de Catalunya Program ICREA Academia (I.P.) and AGAUR Project No.  2021SGR-673 (C.C., I.P.)

\bibliography{references}

\appendix

\section{A -- Asymptotic behavior of colloidal velocity}

The colloidal velocity in the pair force scenario can be written in frequency domain and using dimensionless parameters as 
\begin{equation}
    \utilde(\alpha(\omega)) = \frac{1-(1+\alpha)\bar{v}_0^s(\alpha)-(1/3) \alpha^2 \bar{v}_0^v(\alpha)}{1+\alpha+\alpha^2/\beta}
\end{equation}
where $\utilde(\alpha(\omega))$ is written in units of $\Delta p/\gamma_0$, frequency is given in units of $t_a^{-1}$ and length in units of $a$.
We recall that under these unit convention $\alpha=\sqrt{i\omega}$.
Exploiting the symmetry properties of the real and imaginary components of $\utilde(\alpha)$ we can write the colloidal velocity as 
\begin{equation}
    u(t) = \frac{1}{\pi} \int_0^{\infty} d\omega 
    \left\{ 
        \mathfrak{Re}[\utilde(\alpha)]\cos\omega t -
        \mathfrak{Im}[\utilde(\alpha)]\sin\omega t
    \right\}
\end{equation}
and we may analyze the asymptotic behavior of $\utilde(\omega)$ in the limits $\omega\to 0$ and $\omega\to \infty$ in order to gain insight on the behavior of $u(t)$ for $t\to \infty$ and $t\to 0$, respectively.

\textit{A1 -- Long time behavior} 
Taylor expanding $\utilde(\alpha)$ for small $\alpha$ we obtain 
\begin{equation}
    \utilde(\alpha) = 
    \utilde^{(0)}(d) + \utilde^{(2)}(d,\beta)\alpha^2 + \utilde^{(3)}(d,\beta)\alpha^3 + \Ocal(\alpha^4)
    \label{eq:u.w.expansion}
\end{equation}
where, for small $d\ll1$, we obtain 
\begin{subequations}
    \begin{equation}
        \utilde^{(0)}(d)=\frac{3}{2}d^2+\Ocal(d^3)
    \end{equation}
    \begin{equation}
        \utilde^{(2)}(d,\beta)=-\frac{3}{2}d^2/\beta+\Ocal(d^3)
    \end{equation}
    \begin{equation}
        \utilde^{(3)}(d,\beta)=-\frac{3}{2}d^2/\beta+\Ocal(d^3)
    \end{equation}
\end{subequations}
from which we can find that in the strict limit $\omega=0$ we obtain the total displacement of the colloid as $\Delta r\approx 3d^2/2$ and that in the additional limit $d=0$ the colloid does not move. 

Secondly, one may notice the lack of a linear term in $\alpha$ in eq. \ref{eq:u.w.expansion}, which is present in the absence of the reaction term (\textit{i.e.} if $\bar{v}_0^s=\bar{v}_0^v=0$). 
This linear term in $\alpha$ leads to a dominant real $\omega^{1/2}$ scaling for $\utilde(\omega)$ which is the origin of the $u(t)\sim t^{-3/2}$ long-time tail that is well reported in the literature. 

Thirdly, in contrast, in Eq. \ref{eq:u.w.expansion} the first correction to the $\omega=0$ mode is quadratic in $\alpha^2=i\omega$, indicating that it is purely imaginary. 
This term contributes to $\utilde(\omega)$ on the order of $\Ocal(\omega^2)$, because it is purely imaginary and it is therefore multiplied by $\sin\omega t\sim \omega t$ for small $\omega$. 
Therefore, the first real correction to the zero-order term is the $\alpha^3$ term which had both real and imaginary parts. 
Importantly, this cubic term scales as $\omega^{3/2}$, leading to a $t^{-5/2}$ longtime tail, which arises from the presence of a hydrodynamic dipole in the system. 
Finally, in addition to the temporal scaling we can also observe that the multiplicative coefficient to the cubic term in $\alpha$, $\utilde^{(3)}(d,\beta)\sim 1/\beta$ is inversely proportional to $\beta$, which explains the behavior obtained in Fig. S3.

\textit{A2 -- Short time behavior} 
In the limit of large $\alpha\to \infty$ we may recall the explicit expressions of $\bar{v}_0^s(\alpha)$ and $\bar{v}_0^v(\alpha)$ (respectively Eq. B8 and B9 in the SM), where the hyperbolic functions can be safely approximated as $\sinh\alpha\approx\cosh\alpha\approx e^{\alpha}/2$ to obtain 
\begin{flalign}
\utilde(\alpha) \approx &
    \frac{1}{1+\alpha+\alpha^2/\beta} \times \notag \\ 
    &\left[
    \frac{3 (1+\alpha R)e^{-\alpha d} + \alpha  \left(\alpha  \left(R^3-1\right)-3\right)-3 }{\alpha^2 R^3}
    \right]
\end{flalign}
where $R=1+d$. 
In the additional limit $d\to\infty$ where the reaction takes place infinitely far away from the colloidal surface, multiplicative factor under brackets goes to $1$. 
Note that this is equivalent to making $\bar{v}_0^s=\bar{v}_0^v=0$, indicating that in this regime we recover the expected behavior in the absence of a reaction. 

On the other hand, for finite but small values of $d$, the term under brackets can be split in two contributions: 
one that scales with $e^{-\alpha d}$ and the rest, which allows to define a time from the condition $\alpha d\sim 1 \to t_d\sim d^2$ (or, being explicit with units, $t_d=d^2/\nu$). 
Doing a controlled limit where $\alpha\gg1$, $d\ll1$ and the product $\alpha d>1$ is arbitrary but larger than one, we obtain the following scaling behavior 
\begin{equation}
    \utilde(\alpha)\approx \frac{3}{2} 
    \frac{d^2}{1+3d}
    \frac{\alpha}{1+\alpha+\alpha^2/\beta}
    \label{eq:u.alpha.intermediate}
\end{equation}
which scales as $\alpha^{-1}\sim \omega^{-1/2}$ for large values of $\alpha$ (while maintaining $\alpha d$ arbitrary but larger than one). 
This $\alpha^{-1}\sim \omega^{-1/2}$ leads to a $u(t)\sim t^{-1/2}$ scaling, which is found in Fig. \ref{fig:ucol} (b) for intermediate values of $t$ such that $d^2\ll t \ll 1$.
In frequency this corresponds to $1\ll \alpha \ll d^{-1}$ which indicates that the asymptotic regime found in Eq. \ref{eq:u.alpha.intermediate} corresponds to the $u(t)\sim t^{-1/2}$ shown in Fig. \ref{fig:ucol} (b).

\section{B -- Colloidal velocity for an arbitrary orientation}

We consider a steady dipolar scenario where force densities are 
\begin{equation}
    \FF_{ext}(\rr) = 
    F_0 \pphat 
    \left[ 
        \delta(\rr-\rr_2)
        -\delta(\rr-\rr_1)
    \right]
\end{equation}
with $\rr_1=\RR-\frac{1}{2}\delta \pphat$ and $\rr_2=\RR+\frac{1}{2}\delta \pphat$ with $\RR = -R\zzhat$ and $R = a+d$. 
The unit vector $\pphat$ is aligned with the direction of the dipole $\pphat(\varphi) = \cos\varphi \zzhat+\sin\varphi \xxhat$. 
This allows us to solve the projection of the average velocity in the $Z$ direction   $\langle u (\varphi)\rangle$ which can be approximated, assuming $\delta\ll a$, as 
\begin{equation}
    \langle u (\varphi)\rangle_{T_0} \approx
    \frac{F_0}{4\pi \eta a } 
    \delta 
    \left[
        \cos^2\varphi ~ \frac{\sqrt{R^2-1}}{R^3} - 
        \frac{1}{2}\sin^2\varphi ~ \frac{R^2-1}{R^4}
    \right]
    \label{eq:u.phi}
\end{equation}
where $\delta$ and $R=1+d$ are written in terms of $a$. 
The expression of the colloidal velocity in terms of the angle of the dipole allows to relate to standard scenarios of swimmer propulsion depending on the flow generated around the colloid: 
for $\varphi=0$ the flow direction in relation to the colloidal motion is characteristic of a pusher swimmer while for $\varphi=\pi/2$ it is of a puller one.

\end{document}